\documentclass[11pt]{article}

\usepackage[iso-8859-7]{inputenc}
\usepackage{epsfig}
\usepackage{graphicx}

\usepackage{amsfonts}
\usepackage{amssymb}
\usepackage{amsthm}
\usepackage{amsmath}
\usepackage{multirow}

\newcommand{\newc}{\newcommand}
\newc{\ra}{\rightarrow}
\newc{\lra}{\leftrightarrow}
\newc{\be}{\begin{equation}}
\newc{\ee}{\end{equation}}
\newc{\ba}{\begin{eqnarray}}
\newc{\ea}{\end{eqnarray}}
\newc{\ov}{\overline}
\newc{\D}{\Delta}

\newc{\nn}{\nonumber}

\begin{document}
\thispagestyle{empty}

\vskip 2truecm
\vspace*{3cm}
\begin{center}
{\Huge {\bf
 Family symmetries in F-theory GUTs}}\\
\vspace*{1cm}
{\bf S.F. King$^{1}$,
G.K. Leontaris$^{2,3}$,
G.G. Ross$^{3,4}$}\\
\vspace{4mm}
$^1$ School of Physics and Astronomy, University of Southampton,\\
Southampton, SO17 1BJ, UK\vspace{1mm}\\
$^2$ Physics Department, Theory Division, Ioannina University, \\
GR-45110 Ioannina, Greece\vspace{1mm}\\
$^3$ Department of Physics, CERN Theory Division, \\
CH-1211, Geneva 23, Switzerland\vspace{1mm}\\
$^4$ Rudolf Peierls Centre for Theoretical Physics,\\
University of Oxford, 1 Keble Road, Oxford, OX1 3NP, UK
\end{center}

\vspace*{1cm}
\begin{center}
{\bf Abstract}
\end{center}

\noindent
We discuss  F-theory $SU(5)$ GUTs in which some or all of the quark and lepton families are
assigned to different curves and family symmetry enforces a leading order rank one structure
of the Yukawa matrices. We consider two possibilities for the suppression of baryon and lepton
number violation. The first is based on Flipped $SU(5)$ with gauge group $SU(5)\times U(1)_\chi
 \times SU(4)_\perp$ in which $U(1)_\chi$ plays the role of a generalised matter parity. We
present an example which, after imposing a $Z_2$ monodromy, has a  $U(1)_\perp^2$ family
symmetry. Even in the absence of flux, spontaneous breaking of the family symmetry leads
to viable quark, charged lepton and neutrino masses and mixing.  The second possibility
has an R-parity associated with the symmetry of the underlying compactification manifold
and the flux.  We construct an example of a model with viable masses and mixing angles
based on the gauge group  $SU(5)\times SU(5)_\perp$ with a $U(1)_\perp^3$ family symmetry
after imposing a $Z_2$ monodromy.

\vfill
\newpage

\section{Introduction}

The origin of quark and lepton masses and mixing remains one of the key unanswered
questions in the Standard Model. Recently there has been much interest in the possibility
that the fermion mass structure might emerge from
F-theory~\cite{Donagi:2008ca}-\cite{Blumenhagen:2009up}. Most of the analyses to date have
focused on the possibility that the families belong to a single matter curve and the
fermion mass hierarchy results from the case that the mass matrices  have rank one in the
absence of fluxes~\cite{Heckman:2008qa}-\cite{Marchesano:2009rz}. While this provides a
promising structure it requires that there is only a single intersection of the matter
and Higgs curves in the up down and charged lepton sectors. However explicit calculations
\cite{Hayashi:2009bt} for simple geometries suggest that the number  of intersections
must be even. Although there are ways to recover the rank one starting point,  for
example imposing factorization of the matter curves into irreducible
pieces~\cite{Tatar:2009jk,Heckman:2009mn,Marsano:2009wr,Marsano:2009ym,Hayashi:2010zp},
it does not seem to be the norm with the generic case having, a large number of intersections.

An alternative possibility that can lead to fermion mass hierarchy even for the case of
multiple intersections has been explored by Dudas and Palti~\cite{Dudas:2009hu}. Starting
with the group $SU(5)\times SU(5)_{\perp}$ they explored the possibility that the family
fields belong to {\it different} matter curves. As the fields carry different charges
under the $U(1)$ factors of $SU(5)_{\perp}$ (after identifying the monodromy group) the
latter act as family symmetries. Allowing for spontaneous breaking of these symmetries
can lead to an hierarchical structure for the fermion masses. As we shall discuss in this
case multiple intersections do not disturb the hierarchy. Note that, unlike Dudas and
Palti, we will also consider cases with more than one state on a matter curve.

The survey of all possible monodromies presented in~\cite{Dudas:2009hu} gave rise to
models with promising mass structure but they all suffered from the problem that some
R-parity violating term(s) was not forbidden by the family symmetries and thus the models
had unacceptable levels of baryon and/or lepton number violating processes.  In this
paper we shall discuss how this conclusion can be avoided and illustrate the
possibilities by constructing two models with viable fermion mass matrix structure. The
first model is based on the `flipped' $SU(5)$ group, $SU(5)\times U_\chi(1)$, in which
the $SU(2)$ singlet, charge conjugate down and up quarks belong to the $10$ and $\bar 5$
representations respectively, the opposite assignment to the case of conventional
$SU(5)$. In this case the $U(1)_{\chi}$ acts as a generalised matter parity and
eliminates the leading unwanted baryon and lepton number violating terms. The second
model invokes the R-parity that the authors of \cite{Hayashi:2009bt} argue can arise in
F-theory models through a symmetry of the underlying Calabi-Yau manifold and the flux. In
this case one can build viable models based on the normal $SU(5)$ multiplet assignments.

Of course the ultimate aim is to obtain phenomenologically acceptable quark and lepton
mass matrices. The structure of the quark mass matrices is not completely determined by
the measured quark masses and mixing angles. To a good approximation for the hierarchical
structure that follows from spontaneously broken family symmetries it is the terms on the
diagonal and above the diagonal (assuming left-right convention) in the current quark
basis that are fixed by the quark masses and the Cabbibo Kobayashi Maskawa (CKM) matrix.
The terms below the diagonal (again assuming left-right convention) depend on the rotation
of the right-handed (RH) quark components needed to diagonalise the mass matrix and, due to
the absence of charged gauge bosons coupling to the RH quark sector, we have no constraint
 on it. Assuming a symmetric structure a fit to the available data~\cite{Ross:2007az} has
 the form~\footnote{Reference~\cite{Ross:2007az} also discusses further ambiguities associated
 with the phases and threshold effects.}
\ba
M^{d}&=&
\left(\begin{array}{ccc}
0&-1.9i\epsilon^3&2.3\epsilon^3 e^{-i\pi/3}\\
-1.9i\epsilon^3&\epsilon^2&2.1\epsilon^2\\
2.3\epsilon^3 e^{-i\pi/3} &2.1 \epsilon^2&1
\end{array}
\right)\,m_{b_0}
\label{md}
\\
M^{u}&=&
\left(\begin{array}{ccc}
0&0.4\epsilon^4&0\\
0.4\epsilon^4&0.8\epsilon^3&0\\
0&0&1
\end{array}
\right)\,m_{t_0}
\label{mu}
\ea
where $\epsilon=0.15$. Note that  CKM mixing matrix is unchanged if $M^d$ and $M^u$ are
rotated by the same amount (of course the eigenvalues are unchanged by rotations). This
will be important when we discuss the form of the mass matrices in  the flipped SU(5)
case. The structure of eqs(\ref{md},\ref{mu}) has a texture zero in the (1,1) position
that leads to the prediction~\cite{Gatto:1968ss}
\be
V_{us}(M_X)\approx\left|\sqrt{\frac{m_d}{m_s}}+i\sqrt{\frac{m_u}{m_c}}\,\right|\nn
\ee
that gives an excellent fit to $V_{us}$.

Note also that the magnitude of the $(2,3)$ element of $M^d$ is comparable to the $(2,2)$
element; this is potentially a problem for mass matrices ordered by $U(1)$ symmetries that
typically give $O(\epsilon)$. A non-zero entry in the (1,3) position of $O(\epsilon^3)$
is necessary to avoid the relation $V_{ub}/V_{cb}=\sqrt{m_u/m_c}$.~\footnote{However an
alternative symmetric fit (not considered here) is possible with (1,3) elements of both
$M^u$ and $M^d$ being zero providing one allows for a non-zero (1,1) element in $M^u$
(maintaining a zero (1,1) element of $M^d$)~\cite{Antusch:2009hq}. Such a fit allows
a simple explanation of the right unitarity triangle via a phase sum rule.}

As discussed above the data does not strongly constrain the elements of $M^{u,d}$ below
the diagonal and they are limited only by the constraint that the eigenvalues should
approximately remain the same. The same is true of the $(1,3)$ and $(2,3)$ elements of
$M^u$.

With this brief summary of the desired form of the quark mass matrices we turn to the
structure that can come from F-theory in the case that the mass hierarchy is controlled
by the Abelian symmetries.

\section{Flipped $SU(5)$}

In flipped $SU(5)$~\cite{Barr:1981qv,Antoniadis:1987dx} the chiral matter fields of a
single generation, as in ordinary $SU(5)$, constitute the three components of the
$16\in SO(10)$, ($16=10_{-1}+\bar 5_3+1_{-5}$  under the $SU(5)\times U(1)_{\chi}$
decomposition). However,  the definition of the hypercharge includes a component of the
external $U(1)_{\chi}$ in such a way that flips the positions of $u^c,d^c$ and
$e^c,\nu^c$ within these representations, while leaves the remaining unaltered. Indeed,
employing the hypercharge definition $$Y=\frac 15\left(x+\frac 16 y\right)$$
where, $x$ is the charge under the $U(1)_{\chi}$ and $y$ the `non-flipped' $SU(5)$
hypercharge generator, we obtain the following `flipped' embedding of the SM representations
\ba
F_i&=&10_{-1}\;=\;(Q_i,d^c_i,\nu^c_i)\\
\bar f_i&=&\bar 5_{+3}\;=\;(u^c_i,\ell_i)\label{5M}\\
\ell^c_i&=&1_{-5}\;=\;e^c_i
\ea
In the field theory model the Higgs fields are found in
\ba
H\equiv 10_{-1}\;=\;(Q_H,D_H^c,\nu_H^c)&,&\ov H\equiv \ov{10}_{+1}\;=\;(\bar Q_H,\bar
d_H^c,\bar \nu_H^c)\\
h\equiv 5_{+2}\;=\;(D_h,h_d)&,&\bar h\equiv \bar 5_{-2}\;=\;(\bar D_h,h_u)\label{hhb}
\ea
 When $H,\ov{H}$ acquire non-zero  vacuum expectation values (vevs) along their
neutral components $\langle \nu_H^c\rangle=\langle \bar \nu_H^c\rangle =M_{GUT}$,
they  break the  $SU(5)\times U(1)_{\chi}$ symmetry down to the Standard Model (SM) one.
The breaking of the SM gauge symmetry occurs via vev's of the two fiveplets $h,\bar h$ of
(\ref{hhb}) while the coloured triplets become heavy via the supperpotential terms
$HHh+ \ov{ H}\ov{ H}\ov{ h} \ra \langle \nu_H^c\rangle D_H^cD_h +
\langle \bar\nu_H^c\rangle \bar D_H^c\bar D_h $. In F-theory the breaking of the GUT may
be due to the fluxes rather than fundamental Higgs fields.

Note that matter antifiveplets (\ref{5M}) are completely distinguished from the Higgs
antifiveplets (\ref{hhb}), since they carry different $U(1)_{\chi}$ charges and they
do not contain exactly the same components. As a result $U(1)_{\chi}$ or a discrete factor
of it can be used to forbid the R-parity violating terms. This will be crucial in the F-theory
version of the model that we turn to now.

\subsection{Flipped $SU(5)$ in F-theory}
Our starting point is the sequence
\ba
E_8\supset E_5\{=SO(10)\}\times SU(4)
\ra [SU(5)\times U(1)_\chi]\times SU(4)\ra [SU(5)\times U(1)_\chi]\times U(1)^3
\ea
The adjoint representation of $E_8$ then has the $SO(10)\times SU(4),\;SU(5)\times
U(1)_\chi\times SU(4)$ decomposition given by
\ba
248&\rightarrow&(45,1)+(16,4)+(\ov{16},\ov{4})+(10,6)+(1,15)\nonumber \\
&\rightarrow&
(24,1)_0+(1,15)_{0}+(1,1)_{0}+(1,4)_{-5}+(1,\ov{4})_{5}+(10,4)_{-1}+(10,1)_{4}\nonumber
\\
 &&+(10,\ov{4})_{1}+(10,1)_{-4} +(\ov{5},4)_{3}+(\ov{5},6)_{-2}
+(5,\ov{4})_{-3}+(5,6)_{2}
 \label{decomposition}
\ea
respectively. We further assume that appropriate fluxes exist to induce the required
chirality for the matter fields. At the $SO(10)$ level in particular,  this means that
$\# \,16$ 's$-\#\overline{16}$'s$=3$.

To accommodate the $U(1)_{\chi}$ we see that the monodromies must lie in the $U(1)^3\subset SU(4)$.
There are three possible choices for the monodromy group, namely $S_3$, ${\cal Z}_2\times
{\cal Z}_2$ and ${\cal Z}_2$. The first two cases reduce the number of the available matter curves
 to two. The ${\cal Z}_2$ case gives  three matter curves and only it has the possibility of
distinct localization of the three families. Although the first two cases are not a priori excluded,
in this paper we will consider in detail only the ${\cal Z}_2$ monodromy.

\begin{table}[tbp] \centering%
\begin{tabular}{|c|c|c|}
\hline
\textbf{Field} & \textbf{Representation} & $\mathbf{SU(4)}$
component \\ \hline
$Q_{3},D_{3}^{c},\nu^c_3$ & $10 ^{3}_{-1}$ & $\{t_{1},t_{2}\}$ \\ \hline
$Q_{2},D_{2}^{c},\nu^c_2$ & $10 ^{2}_{-1}$  & $t_{3}$ \\ \hline
$Q_{1},D_{1}^{c},\nu^c_1$ & $10 ^{1}_{-1}$  & $t_{4}$ \\ \hline
$U_{3}^{c},L_3$ & $\overline{5}_{3}^{3}$ & $\{t_{1},t_{2}\}$ \\ \hline
$U_{2}^{c},L_2$ & $\overline{5}_{3}^{1}$ & $t_4$ \\ \hline
$U_{1}^{c},L_1$ & $\overline{5}_{3}^{1}$ & $t_4$ \\ \hline
$l_3^c$&$1^{c_3}_{-5}$&$\{t_1,t_2\}$\\ \hline
$l_2^c$&$1^{c_2}_{-5}$&$t_3$\\ \hline
$l_1^c$&$1^{c_1}_{-5}$&$t_4$\\ \hline
$h_{u}$ & $\overline{5}^{h_1}_{-2}$  & $-t_{1}-t_{2}$ \\ \hline
$h_{d}$ & $5^{h_1}_{2} $ & $-t_{1}-t_{2}$ \\ \hline
$\theta_{ij}$ & $1_0^{ij} $ & $t_{i}-t_{j}$ \\ \hline
$10_H^3$ & $10 ^{3}_{-1}$ & $\{t_{1},t_{2}\}$ \\ \hline
$\ov{10}_H^3$ & $\overline{10}^{3}_{1}$ & $-\{t_{1},t_{2}\}$ \\ \hline
\end{tabular}%
\caption{Field representation content under $SU(5)\times U(1)_{\chi}\times SU(4)_\perp$}
\label{Repsflipped}%
\end{table}%

We label the weights of the $SU(4)$ factor in eq(\ref{decomposition}) by $t_i,i=1,\dots,4$,
with $\sum_{i=1}^4t_i=0$. The ${\cal Z}_2$  monodromy acts on $\{t_1,t_2\}$. The $SU(5)$ matter
representations $F_{1,2,3} \in 10$ belong to $(10,4)_{-1}$.  There are three  matter curves and
we assign one family to each:
\be
10^{(3)}_{-1}\;:\;\{t_1,t_2\},\;
10^{(2)}_{-1}\;:\;\{t_3\},\;
10^{(1)}_{-1}\;:\;\{t_4\}\label{10s}
\ee
The fiveplets, $h, \bar h, \bar f_i$, accommodating the Higgs and matter fields must lie
on a subset of the following curves: The Higgs fiveplet responsible for up quark masses is
in $\bar h\in(\bar 5 ,\bar 6)_{-2}$ so there are four possible Higgs curves
\be
\bar{h}\in \bar 5^{(h_1)}_{-2}:\;\{-t_1-t_2\},\;
                    \bar 5^{(h_2)}_{-2}:\;\{-t_3-t_4\},\;
                    \bar 5^{(h_3)}_{-2}:\;\{-t_1-t_3,-t_2-t_3\},\;
                    \bar 5^{(h_4)}_{-2}:\;\{-t_1-t_4,-t_2-t_4\}
\ee
The down quark Higgs is in $h\in( 5 , 6)_{2}$ and lies on one of  the curves~\footnote{ Since
$\sum_{i=1}^4t_i=0$, we could also label $h$-curves as $5^{(h_1)}_{2}:\{t_3+t_4\}$,
$5^{(h_2)}_{2}:\{t_1+t_2\}$ and so on.}
\be
h\in  5^{(h_1)}_{2}:\;\{-t_1-t_2\},\;
                     5^{(h_2)}_{2}:\;\{-t_3-t_4\},\;
                     5^{(h_3)}_{2}:\;\{-t_1-t_3,-t_2-t_3\},\;
                     5^{(h_4)}_{2}:\;\{-t_1-t_4,-t_2-t_4\}
\ee
The fiveplets accommodating the matter fields  belong to $(\bar 5,4)_{3} $ so there are
three possibilities
\be
\bar f_i\in\bar 5^{(3)}_{3}:\;\{t_1,t_2\},\;
               \bar 5^{(2)}_{3}:\;\{t_3\},\;
               \bar 5^{(1)}_{3}:\;\{t_4\}\label{5bs}
\ee
Charged singlet fields accommodating the right handed electrons belong to $(1,4)_{-5}$
curves
\be
\ell_i^c\in  1_{-5}^{c(3)}:\;\{t_1,t_2\},\;
               1_{-5}^{c(2)}:\;\{t_3\},\;
                 1_{-5}^{c(1)}:\;\{t_4\}\label{1s}
           \ee
The neutral singlets descending from the decomposition of $(1,15)$  lie on the curves $t_i-t_j$
and designated as $\theta_{ij}$ and
\ba
\theta_{ij}=1^{(ij)}_{0}&:&\;\{t_i -t_j\},i\ne j,\; i,j=1,2,3,4
\ea

\subsection{F-ermion Masses}
\subsubsection{Rank-1 structure for the quarks and charged leptons}
As discussed above the $U(1)_{\chi}$ plays the role of an R-symmetry. As we shall see the
Abelian symmetries in the $SU(4)$ factor play the role of family symmetries. We want to
have rank one mass matrices in the absence of family symmetry breaking so it immediately
follows for the down quarks that the down quark Higgs should lie in $5^{(h_1)}$ giving
mass to the third generation through the superpotential coupling $W_{down}=10^{(3)}\cdot
10^{(3)}\cdot 5^{(h_1)}$.

Similarly for the up quarks,  assigning  $\bar f_3$ to $\bar 5^{(3)}$ we must choose the
up quark Higgs to lie on $\bar 5^{(h_1)}$ and the third generation up quark gets mass from
the coupling $10^{(3)}\cdot \bar 5^{(3)}\cdot\bar 5^{(h_1)}$. Turning to the charged
lepton mass matrix we must assign the RH $\tau$-lepton to the $1^{c(3)}$ matter field and it
gets mass from the coupling $1^{(c_3)}\cdot\bar 5^{(3)}\cdot 5^{(h_1)}$. The assignment of
the fields is summarised in Table~\ref{Repsflipped}.

Note that the rank one structure of these mass matrices follows from the $U(1)$ symmetries
and does not require a single intersection of the matter curves with the Higgs curve.

\subsubsection{The light quark masses}

In order to generate masses for the first two generations of quarks and charged leptons
it is necessary to break the family symmetries. This will happen if some of the singlet
(familon) fields  $\theta_{ij}$ develop non-vanishing vevs.  In fact, as discussed in
the Appendix,  two fields, $\theta_{13}$ and $\theta_{14}$, do acquire vevs due to the
Fayet-Iliopoulos (FI) terms~\cite{Fayet:1974jb} associated with the family $U(1)$.
 Allowing for these vevs the down quark mass matrix, which is symmetric as it comes from
 the $10\cdot 10\cdot 5$ coupling, has the form (${\cal O}(1)$ couplings are suppressed)
\ba
M^{d}&=&
\left(\begin{array}{ccc}
\theta_{14}^2&\theta_{13}\theta_{14}&\theta_{14}\\
\theta_{13}\theta_{14}&\theta_{13}^2&\theta_{13}\\
\theta_{14}&\theta_{13}&1
\end{array}
\right)\,m_{b_0}
\ea
Here vevs are understood for the familon fields and  we have suppressed the messenger
mass scale, $M$, associated with the higher dimension operators, {\it i.e.}
$\theta_{13}\equiv \langle\theta_{13}\rangle/M$ etc.
Comparing this with eq(\ref{md}) one sees that the down quark eigenvalues are reproduced
with the choice $\theta_{13}=\epsilon$ and $\theta_{14}=\epsilon^2$.

At this stage we cannot yet determine the CKM matrix as it involves the up quark mass
matrix. The form of the latter requires assignment of the two light generations of
$SU(2)$ singlet up quarks to matter curves. If, as for the $SU(2)$ doublet assignment,
we assign them to different matter curves they have the same weight structure as the
doublets and the form of the up quark mass matrix is the same as for the down quarks.
Unless there are unnatural cancelations  involving the ${\cal O}(1)$ couplings this means
the up quark eigenvalues hierarchy will be similar to that of the down quarks and hence
unacceptable. To avoid this we assign both light generations of $SU(2)$ singlet up quarks
to the {\it same} matter curve $\bar 5^{(1)}$. Then we have
\ba
M^{u}&=&
\left(\begin{array}{ccc}
\lambda_1\theta_{14}^2&\theta_{14}^2&\theta_{14}\\
\lambda_2\theta_{13}\theta_{14}&\theta_{13}\theta_{14}&\theta_{13}\\
\lambda_3\theta_{14}&\theta_{14}&1
\end{array}
\right)\,m_{t_0}
\label{upmatrix}
\ea
In this matrix we have explicitly included the factors $\lambda_{i}$ that determine the
ratios of the $(i,1)$ to $(i,2)$ elements because they play an important role in
generating an acceptable up quark mass matrix. Since we have assigned two families to a
single matter curve, if there is only a single intersection of the matter and Higgs
curves generating each of the entries in the first two columns of the mass matrix, then
the $\lambda_i$s are equal and, by a rotation acting on the first two families of $SU(2)$
singlet up quarks, we can make $\lambda_i=0$.  However, as discussed above, we expect
multiple intersections and in this case the $\lambda_i$s need not be equal and the
rotation can only change them by a common constant $\lambda$. Thus the mass matrix can
have rank three. However, for a large number of intersections or if the intersections are
very close together, we expect $(\lambda_i-\lambda)\ll\lambda$ and so in the rotated
basis we arrive at the form of eq(\ref{upmatrix}) but with small $\lambda_i$s.

With this preamble we can now ask whether the form of eq(\ref{upmatrix}) gives an
acceptable mass matrix. The eigenvalues are in the ratio $1:\theta_{13}
\theta_{14}:\lambda_i \theta_{14}^2=1:\epsilon^3:\lambda_i\epsilon^4$.  Comparing this
with eq(\ref{mu}) we see an acceptable pattern of mass eigenvalues is possible if
$\lambda_i={\cal O}(\epsilon^2)$.

\subsubsection{The CKM matrix}
Finally what about the CKM matrix? Clearly the up and down quark mass matrices are not of
the form given in eqs (\ref{md}) and (\ref{mu}). However a simultaneous rotation of the
up and down quark mass matrices (which leaves the CKM matrix unchanged) can make the
$(1,3)$ and $(2,3)$ elements of $M^u$ and $M^d$ vanish provided the ${\cal O}(1)$
coefficients of these elements in the up and down sectors are equal.  The latter is
expected to be the case if the symmetry at the intersection point of the quark and Higgs
curves is enhanced to $SO(10)$ as is possible since the weight structure of the matter
curves in the up and the down sector involved in the $(i,3)$ Yukawa couplings are the
same. In this case the CKM elements $V_{13}$ and $V_{23}$ (approximately) vanish. However
we know $SO(10)$ must be broken by fluxes so the equality of the $(1,3)$ and $(2,3)$
elements of $M^u$ and $M^d$ can only be approximate. Taking this into account and
performing a common rotation of the up and down quark mass matrices we obtain the form
\ba
M^{d}&=&
\left(\begin{array}{ccc}
\theta_{14}^2&\theta_{13}\theta_{14}&\delta_1 \theta_{14}\\
\theta_{13}\theta_{14}&\theta_{13}^2&\delta_2\theta_{13}\\
\delta_1\theta_{14}&\delta_2\theta_{13}&1
\end{array}
\right)\,m_{b_0}
\ea

\ba
M^{u}&=&
\left(\begin{array}{ccc}
\lambda_1\theta_{14}^2&\theta_{14}^2&0\\
\lambda_2\theta_{13}\theta_{14}&\theta_{13}\theta_{14}&0\\
\lambda_3\theta_{14}&\theta_{14}&1
\end{array}
\right)\,m_{t_0}
\label{upmatrixrot}
\ea
where $\delta_1\approx\delta_2$ takes account of the flux breaking effects. Choosing
$\delta_1\approx\delta_2={\cal O}(\epsilon)$
we obtain the same form as is given in  eqs (\ref{md}) and (\ref{mu}) and hence an
acceptable CKM matrix.

\subsection{The lepton sector}

In flipped $SU(5)$ leptons and down quarks receive masses from couplings not related by
$SU(5)$. Geometrically, RH electrons and down quarks reside on different matter curves.
Thus, in contrast to $SU(5)$, in flipped $SU(5)$ there is no GUT relation between the
 Yukawa couplings of the down quarks and the leptons. However, if we distribute lepton
 doublets to distinct curves as we did for the down quarks, the structure of $M^d$ and
 $M^{\ell}$ will be the same. In this case the situation is similar to that in normal $SU(5)$
and one expects the magnitude of the coefficients to be similar if the geometrical
structure of the relevant intersections  giving rise to the Yukawa couplings in the down
quark and charged lepton sectors are the same. Since the situation is the same as for
ordinary $SU(5)$ we postpone a discussion of how this can lead to an acceptable charged
lepton mass matrix to Section~\ref{lepton}.

Turning to neutrino masses, note that the Dirac neutrino mass matrix originates from the
coupling $10\cdot \bar 5\cdot \bar 5$ and therefore is related to the up quarks. Since the
latter is related to the CKM mixing and has small mixing angles, the large neutrino angles
must be attributed to the see-saw mechanism~\cite{Minkowski:1977sc} and the specific form of
the RH Majorana mass matrix. Doing this is a non-trivial task but may be
 possible~\cite{Smirnov:1993af}. Starting from a near diagonal Dirac neutrino mass matrix
$M^{\nu}_{Dirac}\approx {\mathrm diag}(m_u,m_c,m_t)$ the condition on the heavy
RH Majorana mass matrix $M_R$ in order to yield bi-large neutrino mixing is obtained from the following
 generalization of the string instanton results in~\cite{Antusch:2007jd} to the case of
right-handed neutrinos and arbitrary lepton mixing:
\ba
M_R= \frac{AA^T}{m_1}+\frac{BB^T}{m_2}+\frac{CC^T}{m_3}
\label{MR0}
\ea
where $A=M^{\nu}_{Dirac}\Phi_1$,
$B=M^{\nu}_{Dirac}\Phi_2$, $C=M^{\nu}_{Dirac}\Phi_3$, with $\Phi_i$ being the three columns of the lepton mixing matrix $U=(\Phi_1 \ \Phi_2 \ \Phi_3)$, while $m_i$ are the physical neutrino masses.

We now turn to the question whether it is possible to achieve such right-handed neutrino masses in flipped
$SU(5)$. For this purpose we introduce $\ov{10}_H^3$ additional heavy fields, part of additional vectorlike
pairs, $\ov{10}_H^3,10_H^3$ living on the matter curves. The relevant superpotential couplings needed to
obtaining right-handed neutrino masses are given by (suppressing dimensionless order one coefficients),
\ba
\ov{10}_H^3 (10^3 + \theta_{13}10^2 +\theta_{14}10^1) S_{1,2,3}
\label{seesaw1}
\ea
where $S_{1,2,3}$ are singlet fields, part of the massive string sector with masses
$M_S$.
 After integrating out these  fields we find effective operators
of the form,
\ba
M_R&\sim &
\left(\begin{array}{ccc}
\theta_{14}^2 & \theta_{14}\theta_{13} & \theta_{14}\\
\theta_{14}\theta_{13} &  \theta_{13}^2 &\theta_{13}\\
\theta_{14} &  \theta_{13} & 1\\
\end{array}
\right) \langle\ov{10}_H^3 \ov{10}_H^3\rangle .
\label{MR}
\ea
where we have suppressed not only the dimensionless order one coefficients but also all the dimensional
mass scales of order $M_S$ in the denominators which if reinserted would lead to a rank 3 right-handed
 neutrino mass matrix after the $\ov{10}_{H}^3$ acquires a vacuum expectation value $\langle \ov{10}_{H}^3\rangle
 =\langle\ov{\nu}_H^c\rangle$. Its magnitude fixes the magnitude of the right-handed neutrino masses, the heaviest of which should have an approximate mass
 $\langle\ov{10}_H^3 \ov{10}_H^3\rangle/M_S\sim {\cal O}(10^{14-15})$ GeV in order to get light neutrino masses in the observed range, and this is readily achieved.

Comparing eq.(\ref{MR}) here to the desired form (\ref{MR0}) we see that each of the column vectors $A,B,C$
has the general form $  (\theta_{14}\  \theta_{13}\  1)^T\sim(\epsilon^2 \ \epsilon \ 1)^T$
to be compared to the desired general form $(m_u \  m_c \  m_t)^T\sim(\epsilon^6 \ \epsilon^3 \ 1)^T$. This demonstrates the underlying difficulty in obtaining bi-large mixing in flipped $SU(5)$. It is insensitive to the precise details of the see-saw, following simply from the observation that the field combinations $10^3$, $\theta_{13}10^2$ and $\theta_{14}10^1$ have the same $U(1)_\perp^3$ charges and thus are always generated with the same coefficients. The only way we can see to get bi-large mixing without fine tuning combinations of ${\cal O}(1)$ coefficients is to have strong $SU(5)$ breaking so that the messenger mass, $M_{\nu^c}$, in the $\nu^c$ sector is much greater than the messenger mass $M$ in the quark and charged lepton sector. Then terms proportional to $\theta_{13}/M_{\nu^c}$ can be of order $\epsilon^3$ as required for bi-large mixing provided $M/M_{\nu^c}=\epsilon^2$. Terms involving $\theta_{14}$ require a further suppression and this will be the case if we replace $\theta_{14}/M$ in the quark and charged lepton sector by $\theta_{13}\theta_{34}/M^2$ where $\theta_{34}/M=\epsilon$. Then the term $\theta_{13}\theta_{34}/M_{\nu^c}^2=\epsilon^6$ as required for bi-large mixing (up to the ${\cal O}(1)$ coefficients). While this may be a possible solution to get a viable neutrino mixing pattern it is certainly not very convincing. The price one pays for a viable mass matrix is a complicated choice of vevs and messenger masses; essentially
one exchanges the parameters in the neutrino mass matrix for another set of parameters,
the vevs, and the problem of understanding the neutrino mass matrix structure is replaced
by the problem of determining the vacuum structure of the multi-field familon potential. As we shall discuss the situation is better in the normal $SU(5)$ case where the Dirac neutrino mass matrix is not related to the quark mass matrices.

\subsection{Nucleon decay}

A big advantage of flipped $SU(5)$ is that the $U(1)_\chi$ factor eliminates the
unacceptable dimension 4 baryon- and lepton-number violating operators of the form
$10_M^i\bar 5_M^j\bar 5_M^k$. The symmetry does  however allow baryon and lepton number
operators of dimension five that mediate nucleon decay. They have the form
$10_M^i10_M^j10_M^k\bar 5_M^l$ and their family structure is given by
\ba
{\cal W}_5&\supset&10^3\,10^3\,10^2\,\bar 5^1\,+\,10^3\,10^3\,10^1\,\bar
5^2\,+\,10^3\,10^2\,10^1\,\bar 5^3\nn
\ea
Note that since we have not assigned matter to the $\bar{5}^2$ curve the second operator
is absent. The remaining operators are generated  via heavy triplet mediated graphs and
are expected to be suppressed by the string scale. By itself this is not sufficient
suppression but note that each of the allowed operators involves two matter fields
belonging to the third family of current quarks. This means that the proton decay
operators involving light quarks are further suppressed by small mixing angles and this
can provide the additional suppression needed to bring nucleon decay within experimental
limits.

\section{An $SU(5)$ model}

As pointed out by Hayashi et al~\cite{Hayashi:2009bt} it is possible that the F-theory
has an R-symmetry that descends from a symmetry of  the underlying Calabi-Yau manifold
and the flux. In this case it was shown that there may be both R-parity odd and even zero
modes on a given curve. Assigning the quarks and leptons to odd R-parity states and the
Higgs to even R-parity states,  the leading baryon and lepton number violating
interactions are forbidden even though the $U(1)$s may allow them. This opens up the
possibilities for constructing realistic models based on $SU(5)$ so one must reconsider
the models first analyzed by Dudas and Palti~\cite{Dudas:2009hu}. Here we present  a
model that can closely duplicate the phenomenologically viable mass matrices of eqs
\ref{md} and \ref{mu}.

\subsection{Quark masses}
\begin{table}[tbp] \centering%
\begin{tabular}{|c|c|c|c|}
\hline
\textbf{Field} & \textbf{Representation} & $\mathbf{SU(5)}_{\perp }$ component &
R-parity\\ \hline
$Q_{3},U_{3}^c,l^c_3$ & $\left( 10,5\right) $ & $t_{1,2}$&$ -$ \\ \hline
$Q_{2},U_{2}^{c},l^c_2$ & $\left( 10,5\right) $ & $t_{3}$& $-$ \\ \hline
$Q_{1},U_{1}^{c},l^c_1$ & $\left( 10,5\right) $ & $t_{4}$& $-$ \\ \hline
$D_{3}^{c},L_3$ & $(\overline{5},10)$ & $t_{3}+t_{5}$&$-$ \\ \hline
$D_{2}^{c},L_2$ & $(\overline{5},10)$ & $t_{1}+t_{3}$ &$- $\\ \hline
$D_{1}^{c},L_1$ & $(\overline{5},10)$ & $t_{1}+t_{4}$& $- $\\ \hline
$H_{u}$ & $\left( 5,\overline{10}\right) $ & $-t_{1}-t_{2}$&$ +$ \\ \hline
$H_{d}$ & $\left( \overline{5},10\right) $ & $t_{1}+t_{4}$&$+$\\ \hline
$\theta _{ij}$ & $\left( 1,24\right) $ & $t_{i}-t_{j}$ &$+$\\ \hline
$\theta'_{ij}$ & $\left( 1,24\right) $ & $t_{i}-t_{j}$ &$-$\\ \hline
$S'$ & $\left( 1,1\right) $ & $-$ &$-$\\ \hline

\end{tabular}%
\caption{Field representation content under $SU(5)\times SU(5)_{\perp}$}
\label{Reps}%
\end{table}%

The starting point is the $SU(5)\times SU(5)_{\perp }$ group. The weights of
$SU(5)_\perp$ are labeled by $t_i,\;i=1,\dots, 5$. We will analyse the model  with monodromy
group $Z_{2}$
relating $t_{1}\leftrightarrow t_{2}.$ We assign the quarks and Higgs fields to the
curves as shown in Table \ref{Reps}. In addition there are familon fields $\theta_{ij}$
belonging to the $(1,24)$ representation. With these assignments  the up quark matrix
mass matrix has the form:
\ba
M^{u}/m_{t}=\left(
\begin{array}{ccc}
\theta _{14}^{2} & \theta _{13}\theta _{14} & \theta _{14} \\
\theta _{13}\theta _{14} & \theta _{13}^{2} & \theta _{13} \\
\theta _{14} & \theta _{13} & 1%
\end{array}%
\right)
\ea
where we have written $\theta _{(1,2)j}=\theta _{1j}$ and, for the moment, we allow for
all possible vevs of the familon fields.

The down quark mass matrix has the form:
\ba
M^{d}/m_{b}=\left(
\begin{array}{ccc}
\theta _{54}\theta _{34} &
\theta _{54} & \theta _{14} \\
\theta _{54} & \theta _{53}& \theta _{13}\\
\theta _{31}\theta _{54}+\theta _{34}\theta _{51} & \theta _{51} & 1%
\end{array}%
\right)
\ea
For $\theta_{34}=0$ there is a $(1,1)$ texture zero in the down quark mass matrix. The
choice $\theta_{51}=0$ gives further zeros in the $(3,1)$ and $(3,2)$ positions,
consistent with the data since the elements below the diagonal are poorly determined. To
determine the non-zero familon vevs consider the magnitudes of the quark masses. We
assume that there are no (unnatural) cancelations involving the unknown ${\cal O}(1)$
coefficients in determining the eigenvalues. Then $m_c/m_t=\theta_{13}^2$,
$m_u/m_t=\theta_{14}^2$, $m_s/m_b=\theta_{53}$ and $m_d/m_b=\theta_{54}^2/\theta_{53}$.
The choice $\theta_{53}=\epsilon^2$, $\theta_{54}=\epsilon^3$, $\theta_{13}=3\epsilon^2$,
 $\theta_{14}=\epsilon^3$ and $\theta_{31}=0$ gives a good description of these mass
ratios (up to ${\cal O}(1)$ coefficients) and has the mass matrices
\be
M^{u}/m_{t}=\left(
\begin{array}{ccc}
\epsilon^6 & 3\epsilon^5 & \epsilon^3 \\
3\epsilon^5 &9 \epsilon^4 &3 \epsilon^2\\
\epsilon^3 & 3\epsilon^2 & 1%
\end{array}%
\right) \;\;
M^{d}/m_{b}=\left(
\begin{array}{ccc}
0 &
\epsilon^3& \epsilon^3 \\
\epsilon^3 & \epsilon^2& 3\epsilon^2\\
0& 0 & 1%
\end{array}%
\right)
\label{md1}
\ee

\noindent again up to ${\cal O}(1)$ coefficients.

Of course one must check that this choice is consistent with the familon potential and
this is discussed in the Appendix. Since the theory has three anomalous $U(1)$s  we
expect at least three familon fields should acquire vevs. As discussed in the Appendix,
because the soft SUSY breaking parameters are scale dependent, it may readily happen that
additional familon fields acquire vevs. The important thing to check is that the theory
is F-flat with this choice of vevs and this is demonstrated in the Appendix.

 Turning to the mixing angles one may see that the contribution to $V_{cb}$ from the up
and the down matrices is of the same order and, as discussed above for the case of
flipped $SU(5)$, allowing for some cancelation between them one may readily obtain the
measured value. The same is true for $V_{ub}$. Finally consider the effect of the texture
zero in the $(1,1)$ position of $M^d$. If the symmetry at the intersection points of the
quark and Higgs curves that generate the Yukawa couplings in the $(1,2)$ block is
enhanced to $SO(10)$  the $(1,2)$ couplings will be symmetric as they correspond to the
$SO(10)$ coupling $16\cdot 16\cdot 10$. This with the texture zero gives a down quark
contribution to $V_{us}=\sqrt{m_d/m_s}$. Including the contribution from the up quark
sector   gives $V_{us}=\sqrt{m_d/m_s}+{\cal O}(\sqrt{m_u/m_c})$, again in good agreement with
the measured value.  It is interesting to note that geometry could ensure a further
texture zero in the $(1,1))$ of the up quark mass matrix so that one obtains the full
Gatto-Sartori-Tonin
relation~\cite{Gatto:1968ss}. This happens if there is no intersection of the up quark
and Higgs curves corresponding to the Yukawa coupling in the $(1,1)$ position.

\subsection{Charged lepton  masses}\label{lepton}
There are hints at a stage of Grand Unification coming from the structure of the charged
lepton masses. In particular, after including radiative corrections corresponding to
threshold corrections and the running to low scales, they can be consistent with the mass
relations $m_b=m_{\tau}$ and $Det(M^d)$=$Det(M^{\ell})$ at the GUT
scale~\cite{Ross:2007az,Antusch:2009rx}. In F-theory it is possible to explain the origin
of such relations
provided we assign the LH and charged conjugate RH charged leptons to the same $SU(5)$
representations as the charge conjugate RH down quarks and LH quark doublets respectively
as given in Table \ref{Reps}.
Then the structure of the charged lepton mass matrix will be the same as that of the down
quarks, eq(\ref{md1}), although the ${\cal O}(1)$ coefficients may differ.
However, provided the symmetry at the intersection points of the lepton, Higgs and
familon curves that generate the Yukawa couplings in the $(1,2)$, $(2,1)$ and $(3,3)$
positions is enhanced to $SU(5)$, the ${\cal O}(1)$ coefficients in the down quark mass
matrix will be the same as that for the charged leptons, giving the mass relations
$m_b=m_{\tau}$ and ${\rm Det }(M^d)= {\rm Det}(M^l)$. Of course these relations will have
corrections due to flux breaking but this may be small. However the big problem is to
explain why there is no equivalent relation for the second generation, namely
$m_{\mu}=m_{s}$. Taking account of the radiative corrections, the measured values of the
masses are in gross disagreement with this relation and favour instead  $m_{\mu}\approx 3
m_{s}$~\footnote{but see~\cite{Antusch:2009rx} for more general possibilities.}. In an
$SU(5)$ GUT one may explain the factor of 3 by arranging through additional symmetries
that the $(2,2)$ element involves a coupling to the vacuum expectation value of a $45$
dimensional representation which is proportional to $B-L$~\cite{Georgi:1979df}. As
required this gives a relative enhancement by a factor 3 for the muon compared to the
strange quark. In the case of F-theory this option is not available as, {\it c.f.}
eq(\ref{decomposition}), the $45$ representation of $SU(5)$ are not present. If the
$SU(5)$ were
enhanced to $SO(10)$ then the $45$ representation of
$SO(10)$ could in principle be used in a similar way but since, {\it c.f.}
eq(\ref{decomposition}), it is a
family singlet it cannot selectively couple to the (2,2) element.
However in F-theory a geometrical explanation is possible because the intersection points
of the lepton, Higgs and familon curves that generate the Yukawa couplings in the $(2,2)$
element need not be at an $SO(10)$ enhanced symmetry point relating the strange quark and
muon couplings. In particular if there happens to be a single intersection for the
strange quark and a triple intersection for the muon one expects there to be the required
factor of 3 enhancement for the muon mass.

\subsection{Neutrino masses}

Finally we consider the neutrino masses. The R-parity allows operators quadratic in the
matter fields and so we can construct operators that violate lepton number by 2 units
provided they are invariant under the gauge symmetries. We note that the combinations
$L_1h^u\theta_{14}$ and $L_2h^u\theta_{13}$ are invariant under the gauge symmetries and
so any combination of two of these operators will be allowed. These give rise to a
Majorana mass matrix for  neutrinos given by
\be
M_{Majorana}^{\nu}=\left(
\begin{array}{ccc}
9\epsilon^4 & 3\epsilon^5 &0 \\
3\epsilon^5 & \epsilon^6 &0\\
0 & 0 & 1%
\end{array}%
\right) \frac{\left(h^u\right)^2}{M}\;\;
\label{md1}
\ee
For the messenger scale $M$ at the string scale $M\gg 10^{10}GeV$ and these masses are
negligible. This means there should be light messengers and the obvious possibility is
that there are light right-handed neutrinos. The R-parity odd  $SU(5)$ singlet fields
$\theta'_{ij}$ and $S'$ are candidate right-handed neutrinos.

A choice that can
accommodate the observed neutrino masses starts with the odd R-parity zero modes
$\theta'_{15}$ and $S'$. Through the superpotential coupling $\lambda SS'^2$ the field
$S'$ acquires a Majorana mass, $M_S'=\lambda S$, if the R-parity even field $S$ acquires
a vev.  As shown in the Appendix F-flatness requires that $\theta_{51}$ also acquires a
vev of ${\cal O}(S\frac{\theta_{53}}{\theta_{13}})$ and this in turn generates a Majorana mass,
$M_{15}$ for $\theta_{15}'$, $M_{15}={\cal O}(\lambda'^2\theta_{51}^2/M_S)$ through the coupling
$\lambda' S'\theta_{15}'\theta_{51}$, assuming a hierarchy $M_{15}\ll M_{S'}$. With such a hierarchy
 the right-handed neutrinos $\theta'_{15}$ and $S'$ have suppressed mixing and  we may apply
the conditions of sequential dominance~\cite{King:1999mb,Antusch:2010tf} to achieve
a neutrino mass hierarchy with large atmospheric and solar mixing as discussed below.

Now the coupling of the LH-neutrino states to $\theta'_{15}$ and $S$ will generate
Majorana masses for two combinations of the LH neutrino states. The dominant term
generating the heaviest (atmospheric) neutrino mass involves the lightest RH neutrino
state, $\theta_{15}'$. Its coupling to the light neutrinos is through the term
(suppressing the ${\cal O}(1)$ coefficients)
$
(L_3\theta_{13}+L_2\theta_{53}+L_1\theta_{54})\theta_{15}'h^u
$
and, through the see-saw mechanism generates the neutrino mass term
\be
(L_3\theta_{13}+L_2\theta_{53}+L_1\theta_{54})^2\langle h^u\rangle^2/M_{15}
\ee
In the fit to the quark masses quoted above we had $\theta_{13}=3\epsilon^2$,
$\theta_{53}=\epsilon^2,\theta_{14}=\epsilon^3$ and $\theta_{54}=\epsilon^3$. This does
not give the observed atmospheric neutrino mixing angles unless the $O(1)$ coefficients
play a role. As a simple example of this we suppose that the coefficient of the $(2,3)$
entry of $M^u$ has a relative factor of 3 in its coupling (as mentioned above this could
readily happen if there are three intersections generating the coupling).  Then the fit
to $M^u$ gives  $\theta_{13}=\epsilon^2$, $\theta_{53}=\epsilon^2$,
$\theta_{54}=\epsilon^3$ and $\theta_{14}=3\epsilon^3$. In this case, up to ${\cal O}(1)$
coefficients, we have the atmospheric neutrino mass term given by
\be
m_@(\nu_{\tau}+\nu_{\mu}+\epsilon\nu_{e})
\ee
where $m_@=\epsilon^4\langle H^u\rangle ^2/M_{15}$. To ${\cal O}(\epsilon)$
one obtains near-maximal atmospheric mixing in
agreement with the observed value.

A second Majorana mass is generated through the see-saw mechanism via the coupling
$(L_2\theta_{13}+L_1\theta_{14})S'h^u\approx
((-\nu_\tau+\nu_\mu)\theta_{13}/2+\nu_e\theta_{14})S'h^u$ where we have kept only the
components left light by the dominant first Majorana mass term. This gives the second
neutrino mass term
\be
m_{\odot} (-\nu_\tau+\nu_\mu+6\epsilon\nu_e)^2
\ee
where $m_{\odot}=\epsilon^4\langle h^u\rangle^2/ (4m_S)$. Since $6\epsilon\approx 0.9$ this gives large
solar mixing. The absolute value of the neutrino masses requires that $S={\cal O}(\epsilon^9)$ corresponding
to Majorana masses for the RH-neutrinos of ${\cal O}(10^{10}GeV)$. The ratio of the solar to
atmospheric masses is of ${\cal O}(1/4)$ up to the ${\cal O}(1)$ factors. Our analysis assumes
$M_{15}<M_S'$ and this can be justified with a reasonable choice of the ${\cal O}(1)$ factors
since  several of these factors are involved. The estimates above of the bi-large mixing pattern are only
valid up to ${\cal O}(\epsilon)$ corrections and further (small) corrections from the charged lepton sector.

A final comment is in order. The assumption that there are light singlet fields $S$ and $S'$ can be questioned as they do not couple to fluxes and so fluxes cannot ensure their chirality. An alternative is to replace $S$ and $S'$ by $\theta_{31}$ and $\theta_{31}'$. Then with $\theta_{31}=\epsilon^7$ one generates a singlet vev for $\theta_{31}\theta_{13}$ of the required order. Similarly we can replace $S'$ by $\theta_{13}\theta_{31}'$. One may readily check that the structure                                                                                                                             of the light neutrino masses and mixing remains the same.

\section{Doublet triplet splitting, the $\mu$ term and FCNC}

 So far, we have discussed how the above GUT models are capable of reproducing the fermion mass hierarchy
  and the CKM mixing.   However it is also necessary to inhibit nucleon decay by making  the colour triplets of the
  fiveplet Higgs fields $h,\bar h$ heavy.  In the Flipped $SU(5)$  model we have already argued that in the presence of  Higgs tenplets $H, \ov{H}$ , there is a doublet-triplet  splitting mechanism and  triplets acquire a mass due to the missing partner mechanism. In the normal $SU(5)$ case this solution is not possible. It has been suggested that the splitting can be achieved by putting the up and down Higgs on different matter curves. As a result there is no direct mass term inducing a dimension-five proton decay operator, whilst heavy mass terms for the triplets are generated when combined with the heavy KK-modes~\cite{Beasley:2008kw}. However it was shown in \cite{Dudas:2009hu} that this solution is not available in the case that the matter fields reside on different matter curves. Given this we must assume  that the geometry accommodates Wilson line breaking in which case it is possible to project out the light triplet states.

   It is also necessary to have a mechanism to generate the $\mu$-term.   For the case that the up and down Higgs curves intersect each other,  a $\mu$ term can be naturally generated through their interaction with a chiral superfield localised on a curve normal
   to the GUT surface~\cite{Beasley:2008kw}.

   Finally we consider the bounds on family symmetries imposed by requiring consistency with the measurements sensitive to flavour changing neutral currents (FCNC). In supersymmetric models the limits on FCNC give rise to stringent bounds on dimension 2 and 3 soft supersymmetry breaking terms \cite{Gabbiani:1988rb}. The latter are very dependent on the precise origin of supersymmetry breaking and can be suppressed in specific schemes so we concentrate here on the former. Of these the strongest bound in the squark sector is on the left-handed $\Delta^{LL}_{ds}\phi_{dL}^\dagger \phi_{sL}$ and right-handed $\Delta^{RR}_{ds}\phi_{dR}^\dagger \phi_{sR}$ soft mass terms mixing the first two generations~\footnote{  For an updated summary of results and extensive references see~\cite{Isidori:2010gz}.}. For gaugino and squarks of comparable order and allowing for the running between the mediator scale and the SUSY breaking scale \cite{de Medeiros Varzielas:2006ma} the most stringent experimental bounds are $\Delta^{LL}_{ds}/\tilde{m}^2<{\cal O}(\epsilon)$ and $\sqrt{\Delta^{LL}_{ds}\Delta^{RR}_{ds}}/\tilde{m}^2<{\cal O}(\epsilon^3)$, where $\tilde{m}^2$ is the mean squark mass squared taken here to be $(350{\rm GeV})^2$.  Both the models discussed here $\phi_{dL,R}^\dagger \phi_{sL,R}$ have weight structure $t_4-t_3$ and the associated mass terms will arise at ${\cal O}(\theta_{31}\theta^{\dagger}_{14})$. In the flipped $SU(5)$ case these terms are of ${\cal O}(\epsilon^3)$ while in the normal $SU(5)$ case it is of ${\cal O}(3\epsilon^5)$, both consistent with the bounds. In gauge family symmetry models there is a second source of these terms coming from the $D-$terms of the family symmetry. On rotating to the down quark mass eigenstate basis these induce the off-diagonal $d-s$ mixing terms.  The  D-terms are proportional to the familon soft mass squared masses \cite{Kawamura:1994ys, de Medeiros Varzielas:2006ma} and if these are of the same order as the mean squark mass  the contribution is of ${\cal O}(\epsilon)$, violating the bounds. Allowing for mean squark masses to be of ${\cal O}(1)$ TeV only reduces the experimental bound by a factor $\epsilon$ so it is necessary that the familon soft masses should be somewhat smaller than the squark masses, a factor of $\epsilon$ being consistent with  a $(350{\rm GeV})^2$ mean squark mass. This may readily happen if the SUSY breaking messenger fields are more weakly coupled to the familons than the squarks.

   These estimates  readily extend to the slepton sector. In this case the predicted value of the $\mu -e$ mixing terms at the messenger mass scale is reduced by approximately $1/3$ because $m_d/m_\mu\approx1/3$ at that scale giving a reduction in the mixing angle needed to diagonalise the lepton mass matrix. The experimental bounds on $\Delta_{e \mu}$ and $\Delta_{ds}$ are comparable and so the overall bound on the familon soft mass coming from the slepton sector is somewhat weaker than that coming from the squark sector.

\section{Summary and Conclusions}

In this work we have presented two examples of viable fermion mass textures of quarks charged leptons
and neutrinos in the context of local F-theory GUTs. In these models the fermion mass hierarchy is ensured by family symmetries and spontaneous breaking of these symmetries can give viable masses and mixings even in the absence of flux corrections.

The first example is based on the Flipped  $SU(5)\times U(1)_{\chi}$ gauge symmetry in which the fermion generations
carry charges under the two Abelian factors of the enhanced (family) gauge symmetry $U(1)_\perp^2$, left after imposing a ${\cal Z}_2$ monodromy relating two Abelian factors of  $SU(4)_{\perp}$.  A fermion mass pattern consistent with the low energy data arises when matter assigned in $10$'s resides on different matter curves and matter transforming under
$\bar 5$ is accommodated only in two matter curves. Furthermore, it is shown that  $ U(1)_{\chi}$ acts
as a generalised matter parity, preventing all dangerous $R$ parity breaking (dimension-four) operators. While it may be possible to accommodate a viable pattern of neutrino masses and mixings it must be admitted the resulting structure looks very contrived.

The second example is based on the $SU(5)$ GUT gauge symmetry with matter transforming under the family symmetry $U(1)_\perp^3\subset SU(5)_{\perp}$, while again a  $Z_2$ monodromy is imposed among two $U(1)_\perp$ factors of $SU(5)_{\perp}$. Invoking an $R$-parity that can arise in certain Calabi-Yau compactifications with appropriate fluxes, we construct an $R$-parity conserving model capable of generating the observed quark and lepton masses and mixing angles. In contrast to the previous example, each fermion family is localised  on  a different matter curve. Giving vevs to only a few familon fields we break the $U(1)_\perp$ family symmetries and generate charged fermion mass matrices with the required hierarchy of masses and mixing angles. In addition, using parity-odd singlet fields for right-handed neutrinos, and mildly  extending the singlet (familon) field content that acquire vevs  along F- and D-flat directions, we demonstrate how to construct an effective light neutrino Majorana mass matrix with bi-large mixing and mass squared differences in the experimentally required region.

{\bf Acknowledgements}

We would like to thank Bruce Campbell, Emilian Dudas, Fernando Marchesano, Eran Palti, Pierre Ramond, Sakura Schafer-Nameki and Timo Weigand
for useful discussions.

This work is partially supported by the European Research and Training
Network  (RTN) grant `Unification in the LHC era' (PITN-GA-2009-237920).
SFK acknowledges support from the STFC Rolling Grant ST/G000557/1
and is grateful to the Royal Society for a Leverhulme Trust Senior Research Fellowship.

\newpage

\noindent{\bf{\Large Appendix}}

\noindent{\large {\bf The familon potential in flipped $SU(5)$}}

The superpotential terms involving the familon fields $\theta_{ij}$ is
\ba
{\cal W}_{\theta}&=&\lambda_{ijk}\theta_{ij}\theta_{jk}\theta_{ki}\nn\\
&=&\lambda'_{1}\theta_{13}\theta_{34}\theta_{41}+\lambda'_{2}\theta_{31}\theta_{14}\theta_{43}
\ea

If only $\theta_{ij}$ acquire vevs at a high scale, the flatness conditions read
\ba
\frac{\partial{\cal
W}_{\theta}}{\partial\theta_{ij}}&=&\lambda_{ijk}\theta_{jk}\theta_{ki}=0
\label{FC}
\ea
For our choice of non-zero vevs ($\langle\theta_{13}\rangle\ne
0,\langle\theta_{14}\rangle\ne 0$) conditions (\ref{FC})
 are automatically satisfied.

 To write down the corresponding $D$-flatness conditions, we must take into account the
monodromies.
 For  the ${\cal Z}_2$ monodromy, $t_1\leftrightarrow t _2$  the $D$-flatness conditions
can be written in compact form
\ba
\sum_{j=3,4}\left|\langle\theta_{nj}\rangle\right|^2-\left|\langle\theta_{jn}\rangle\right|^2+\xi_n&=&0,\;n=3,4
\ea
where $\xi_n$ are -moduli dependent- FI terms. For the specific choice of vevs these read,
\ba
-\left|\langle\theta_{13}\rangle\right|^2+\xi_3&=&0\nn\\
-\left|\langle\theta_{14}\rangle\right|^2+\xi_4&=&0\nn
\ea
Note that these equations require two familon fields acquire vevs and these must be
$\theta_{13}$ and $\theta_{14}$ if  $\xi_3$ and $\xi_4$ are positive.

In the presence of large vevs for  possible $H_i=10_i, \ov{H}_i=\ov{10}_i$ Higgs
 fields, the D-flatness conditions are modified as follows
 \ba
10 \left(\left|\langle H_n\rangle\right|^2- \left|\langle\ov{H}_n\rangle\right|^2\right)+
\sum_{j=1,3,4}\left|\langle\theta_{nj}\rangle\right|^2-\left|\langle\theta_{jn}\rangle\right|^2+\xi_n&=&0,\;n=3,4
 \ea
 and an analogous solution can be worked out.

 \noindent{\large {\bf The familon potential in $SU(5)$}}

 In this case there are twelve familon fields of the form $\theta_{ij}\;i,j=1,3,4,5$ and
three $U(1)$s. This means we expect at least three vevs for the familon fields to be
required by the D-flatness condition. To generate the quark and charged lepton masses we
require vevs for four fields  $\theta_{53},\;\theta_{54},\;\theta_{13}$ and $\theta_{14}$
and so we must check that it is possible for more than three familons to get vevs. From
eq(\ref{FC}) we see that the choice of vevs is F-flat. The D-flatness conditions are
 \ba
-\left|\langle\theta_{13}\rangle\right|^2-\left|\langle\theta_{53}\rangle\right|^2+\xi_3&=&0\nn\\
-\left|\langle\theta_{14}\rangle\right|^2-\left|\langle\theta_{54}\rangle\right|^2+\xi_4&=&0\nn\\
\left|\langle\theta_{53}\rangle\right|^2+\left|\langle\theta_{54}\rangle\right|^2+\xi_5&=&0\nn
\label{DT}
\ea
and clearly can be satisfied for $\xi_5$ negative and $\xi_3,\;\xi_4$ positive. However
these equations have a flat direction corresponding to the fact that we require four
familon vevs but there are only three D-terms.  The familon potential also has soft SUSY
breaking mass terms. If these are constant then only the three familon fields with the
smallest (positive) mass squared will acquire vevs. However the mass squared terms are
scale dependent due to the Yukawa couplings that increase the soft mass squared as the
scale is increased. Thus the contribution to the potential of the soft mass squared terms
has the form
\be
V(\theta_{ij})=m_{13}^2(\phi_{13})\left|\langle\theta_{13}\rangle\right|^2+m_{14}^2(\phi_{14})
\left|\langle\theta_{14}\rangle\right|^2+m_{53}^2(\phi_{53})\left|\langle\theta_{53}
\rangle\right|^2+m_{54}^2(\phi_{54})\left|\langle\theta_{54}\rangle\right|^2
\label{ST}
\ee
Minimising eqs(\ref{DT}) and (\ref{ST}) can readily require all four vevs to be non-zero.

The discussion has so far dealt with the vevs required to give the quarks and charged
leptons a mass. However in order to generate a mass for the neutrinos further (much
smaller) vevs were needed. Consider the case where the additional vevs are for the fields
$S$ and $\theta_{51}$. In this case the F-term conditions may change due to the additional
couplings of the form $S\theta_{ij}\theta_{ji}$. If only the fields acquiring vevs are light
no additional F-terms appear. If however the field $\theta_{35}$ is also light we have
non-trivial term given by
\ba
\left|\langle F_{35}\rangle\right|^2=
\left|\langle\theta_{13}\theta_{51}+S\theta_{53}\rangle\right|^2\nn
\ea
This requires $S=O(\theta_{51}\frac{\theta_{13}}{\theta_{53}})$. The D-term conditions can
 be satisfied with only very
small changes in the dominant vevs because they are quadratic in the fields. This changes
the F-terms (linear in the fields)  by small corrections and they can be compensated by
small corrections to the $S$ and $\theta_{51}$. Repeating the procedure one obtains a
rapidly convergent perturbative solution to the D- and F-flatness conditions. No additional
 non-trivial F-terms are generated in the case that $S$ is replaced by the field $\theta_{31}$
  that acquires a vev.

 \end{document}